\documentclass[aps,prb,twocolumn,showpacs,showkeys,groupedaddress,floatfix,amsmath,amssymb]{revtex4}
\usepackage{graphicx}
\usepackage{dcolumn}
\usepackage{bm}

\begin{document}


\title{Multiple flux jumps and irreversible behavior of thin Al superconducting
rings}

\author{D. Y. Vodolazov}
\author{F. M. Peeters}
\email{peeters@uia.ua.ac.be}
\affiliation{Departement Natuurkunde,
Universiteit Antwerpen (UIA), Universiteitsplein 1,  B-2610
Antwerpen, Belgium}

\author{S. V. Dubonos}
\altaffiliation[Also at: ] {Institute of Microelectronic
Technology, Russian Academy of Sciences, Chernogolovka, 142432,
Russia}
\author{A. K. Geim}
\email{geim@man.ac.uk} \affiliation{Department of Physics and
Astronomy, University of Manchester, Oxford Road, M13 9PL,
Manchester, United Kingdom}

\date{\today}

\begin{abstract}

An experimental and theoretical investigation was made of flux
jumps and irreversible magnetization curves of mesoscopic Al
superconducting rings. In the small magnetic field region the
change of vorticity with magnetic field can be larger than unity.
This behavior is connected with the existence of several
metastable states of different vorticity. The intentional
introduction of a defect in the ring has a large effect on the
size of the flux jumps. Calculations based on the time-dependent
Ginzburg-Landau model allows us to explain the experimental
results semi-quantitatively.
\end{abstract}

\pacs{74.60Ec, 74.20.De, 73.23.-b}

\maketitle

\section{Introduction}

Recently, Pedersen {\it et al.} \cite{Pedersen} observed jumps in
the magnetization of superconducting rings which corresponds to
changes of the vorticity larger than unity. This is in contrast to
the behavior of superconducting disks where only changes in the
value of the vorticity of unit size were observed \cite{GeimN}. In
some respect the observed behavior in rings is similar to vortex
avalanches which were observed in superconductors with strong bulk
pinning \cite{Zieve,Behina} or to jumps in the magnetization when
several vortices (in the form of a chain) enter in a
superconducting film of width comparable to the coherence length
\cite{Bolech,Vodolazov}. The occurrence of such jumps in a defect
free superconducting ring originates from the fact that several
metastable states with different vorticity $L$ are possible for a
given magnetic field. However the existence of such multiple
stable states is not a sufficient condition to explain changes in
the vorticity larger than unity (e.g. they also exist in the case
of superconducting disks). An additional important requirement is
to find the stability condition and finally the state to which the
system relaxes to. This requires the study of the transition
process from one state to another, i.e. it requires the analysing
the time-dependent process.

The stability condition was studied numerically in Ref.
\cite{Fink} for the case of a hollow cylinder, and in a number of
works (see for example Ref. \cite{Baelus,Berger} and references
therein) for superconducting disks and rings by using the static
Ginzburg-Landau (GL) equations. Unfortunately no analytical
results were presented due to the rather general character of the
studied systems in the above works.

Recently, we studied the transition process \cite{our1} using the
time-dependent Ginzburg-Landau equations. It was shown that
transitions between different metastable states in a mesoscopic
superconducting ring are governed by the ratio between the time
relaxation of the phase of the order parameter $\tau_{\phi}$
(which is inversely proportional to the Josephson frequency) and
the time relaxation of the absolute value of the order parameter
$\tau_{|\psi|}$. We found that if the ratio
$\tau_{|\psi|}/\tau_{\phi}$ is sufficiently large the system will
always transit from a metastable state to the ground state. This
leads to an avalanche-type variation of $L$ when the vorticity of
the metastable state differs appreciably from the vorticity of the
ground state. In contrast to the case of a superconducting film,
in a ring the 'vortex' entry occurs through a single point and the
vorticity increases one by one during the transition. In
low-temperature superconductors like In, Al, Sn the ratio
$\tau_{|\psi|}/\tau_{\phi}$ is very large for temperatures far
below the critical temperature $T_c$ and hence, if such systems
are driven far out of equilibrium they will always relax to the
ground state.

In this work we investigate the conditions under which a state
with a given vorticity becomes unstable in a finite width ring and
we find how the supeconducting order parameter in the ring changes
with increasing applied magnetic field. We are able to find an
analytical expression for the dependence of the order parameter on
applied magnetic field, and hence for the upper critical field at
which superconductivity vanishes in such a sample. We provide a
direct comparison between the theoretical and experimental results
on aluminium rings. Our theoretical calculations are based on a
numerical solution of the time-dependent Ginzburg-Landau
equations.

The paper is organized as following. In section II the theoretical
formalism is presented and the two-dimensional time-dependent GL
equations are solved. In Section III the experimental results are
presented and compared with our theory. In Section IV we present
our conclusions and our main results.

\section{Theory}

We consider sufficient narrow rings such that we can neglect the
screening effects. This is allowed when the width of the ring $w$
is less than ${\it max}{\lambda,\lambda^2/d}$, where $\lambda$ is
the London penetration length and $d$ is the thickness of the
ring. In order to study the response of such a ring on the applied
magnetic field we use the time-dependent Ginzburg-Landau equations
\begin{subequations}
\begin{eqnarray}
u \left(\frac {\partial \psi}{\partial t}+i\varphi\psi  
\right) & = & (\nabla - {\rm i} {\bf A})^2 \psi +(1-|\psi|^2)\psi,
\quad
\\
\Delta \varphi & = &  {\rm div}\left({\rm Im}(\psi^*(\nabla-{\rm
i}{\bf A})\psi)\right), \quad 
\end{eqnarray}
\end{subequations}
where all the physical quantities (order parameter
$\psi=|\psi|e^{i\phi}$, vector potential $A$ and electrostatical
potential $\varphi$) are measured in dimensionless units:  the
vector potential $A$ is scaled in units $\Phi_0/(2\pi\xi)$ (where
$\Phi_0$ is the quantum of magnetic flux), and the coordinates are
in units of the coherence length $\xi(T)$. In these units the
magnetic field is scaled by $H_{c2}$ and the current density, $j$,
by $j_0=c\Phi_0/8\pi^2\lambda^2\xi$. Time is scaled in units of
the Ginzburg-Landau relaxation time $\tau_{GL}=4\pi\sigma_n
\lambda^2/c^2 $, the electrostatic potential, $\varphi$, is in
units of $c\Phi_0/8 \pi^2 \xi \lambda \sigma_n$ ($\sigma_n $ is
the normal-state conductivity). Here the time-derivative is
explicitly included which allows us to determine the moment at
which the state with given vorticity $L$ becomes unstable. It is
essential to include the electrostatic potential (which is
responsible for the appearance of the Josephson time or frequency)
in order to take into account the multi-vortex jumps. In some
previous studies (see for example Refs. \cite{McCumber,Tarlie})
$\varphi=0$ was assumed and as a consequence only transitions with
unit vorticity jumps, i.e. $\Delta L=1$, are possible in the ring
\cite{our1}. The coefficient $u=48$ was chosen such that after the
transition the system is in the thermodynamically equilibrium
state \cite{our1}. We assume that the width ($w$) of the ring is
less than two coherence length $\xi$, because: i) all experimental
results presented here were performed for such type of samples;
and ii) only in this case it is possible to obtain simple
analytical expressions. For instance, the dependence of the order
parameter on the applied magnetic field and the upper critical
field $H_{max}$.

For $w\leq 2\xi$ the order parameter is practically independent of
the radial coordinate. This is demonstrated in Fig. 1 where the
dependence of the order parameter in the middle of the ring is
compared with its value  at the inner and outer boundary of the
ring, i.e. $r=R\pm w/2$ ($R$ is the mean radii of the ring), for
two different rings. Notice that these two numerical examples
corresponds already to relative thick mesoscopic rings, i.e. $R/w
\sim 1-2$. For the field $H_{max}$ we are able to fit our
numerical results to the expression
\begin{equation} 
H_{max}=3.67\frac{\Phi_0}{2\pi \xi w}.
\end{equation}
For rings with $w\leq 2\xi$ and $w/R<1$ this analytical expression
is within $2\%$ of the numerical results. It is interesting to
note that $H_{max}$ does not depend on the radii of the ring. But
the value of the vorticity of the system depends on $R$. For
example, for $R=5.5\xi(16.5\xi)$ $L=55(501)$ for $w=\xi$ at
$H=H_{max}$.

\begin{figure}[h]
\includegraphics[width=0.5\textwidth]{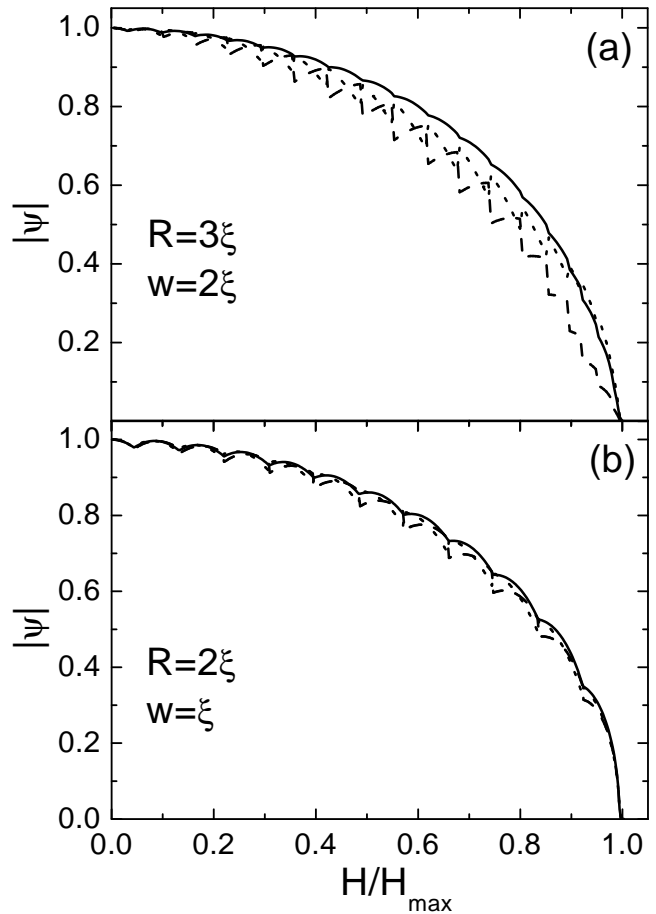}
\caption{Dependence of the absolute value of the order parameter
$|\psi|$ on the applied magnetic field for two different rings in
the ground state. Dashed curve corresponds to $|\psi|(R-w/2,H)$,
solid curve to $|\psi|(R,H)$ and dotted curve to $|\psi|(R+w/2)$.}
\end{figure}

Note that Eq. (2) has the same dependence on the superconducting
parameters as in the case of a thin plate with thickness
$d<\sqrt{5}\lambda$ placed in a parallel magnetic field
\cite{Ginzburg,Douglass}. Even the numerical coefficient is quite
close, i.e. for a thin plate it is equal to $2\sqrt{3}\simeq
3.46$. Furthermore, we found that the transition to the normal
state of our rings at the critical field $H_{max}$ is of second
order as is also the case for a thin plate. A possible reason for
this close similarity is that for a thin plate with thickness
$d<\sqrt{5}\lambda$ the screening effects are also very small. In
the calculations of Refs. \cite{Ginzburg,Douglass} an averaged
value for the order parameter was used independent of the
coordinate. Note that this is similar to our $|\psi|$ which is
practically independent on the radial coordinate (see Fig. 1).

The absolute value of the order parameter (in the middle of the
ring) is, too a high accuracy, given by the expression
\begin{equation} 
|\psi|^2=1-(H/H_{max})^2-p(L,H)^2,
\end{equation}
with $p(L,H)=L/R-HR/2$, where the vorticity $L$ depends on the
history of the system. This result is similar to the one obtained
in Refs. \cite{Ginzburg,Douglass} with the exception of the last
term in Eq. (3) which appears due to the closed geometry of the
ring and hence leads to a nonzero $L$.
\begin{figure}[h]
\includegraphics[width=0.5\textwidth]{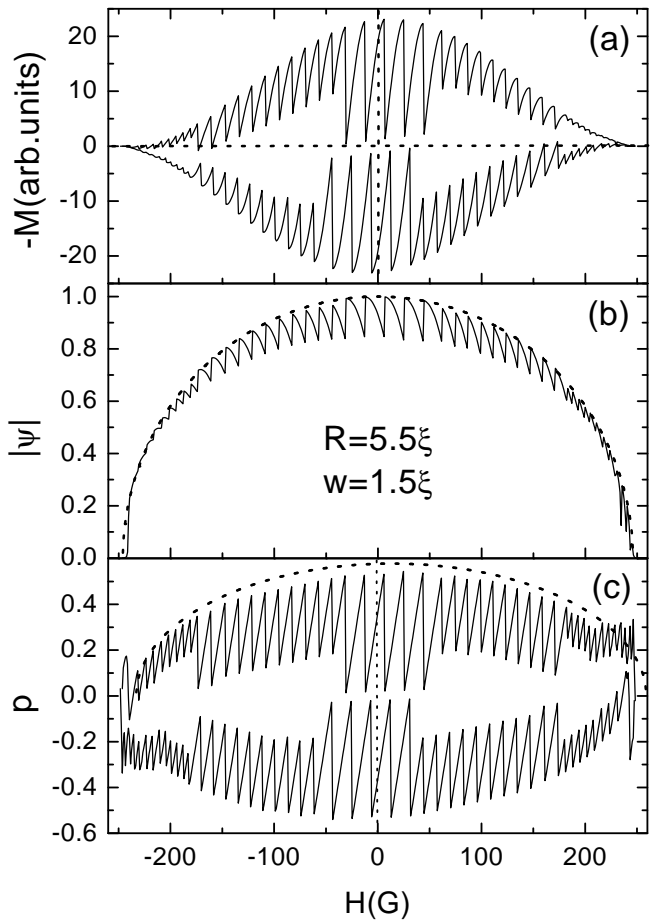}
\caption{Magnetic field dependence of the magnetization (a), the
order parameter (b) and the gauge invariant momentum (c) in the
middle of the ring. Dotted curve in Fig. 2(b), is the expression
$\sqrt{1-(H/H_{max})^2}$. Dotted curve in Fig. 2(c) is the
expression $\sqrt{1-((H-H_0)/H_{max})^2}/\sqrt{3}$, where $H_0
\simeq 13 G $ - is the displacement of the maximum of $M(H)$ from
the $H=0$ line.}
\end{figure}

All the above results were obtained for a ring which is in the
ground state at any value of the magnetic field. However such a
system can exhibit several metastable states at a given magnetic
field, and consequently this may lead to hysteretic behavior when
one sweeps the magnetic field up and down. Furthermore, with
changing field the vorticity may jump with $\Delta L>1$. An
important question which arises is the condition of stability of
the state with given vorticity. This question was studied earlier
for one-dimensional rings \cite{Tarlie,our2}, i.e. rings with zero
width. It turns out that the system transits to a state with
another vorticity when the value of the gauge-invariant momentum
${\bf p}=\nabla \phi -{\bf A}$ reaches the critical value
\begin{equation} 
p_c=\frac{1}{\sqrt{3}}\sqrt{1+\frac{1}{2R^2}}.
\end{equation}
At this condition it is easy to find the value of the field for
the first 'vortex' entry
\begin{equation} 
H_{en}/H_{c2}=2p_c/R=\frac{2}{\sqrt{3}R} \sqrt{1+\frac{1}{2R^2}}.
\end{equation}

We will now generalize the results of Refs. \cite{Tarlie,our2} to
the case of finite width rings with $w\lesssim 2\xi$. First we
will neglect the dependence of $\psi$ on the radial coordinate in
which case the GL equations reduce to one-dimensional expressions.
But in order to include the suppression of the order parameter by
an external field for a finite width ring we add the term
$-(H/H_{max})^2\psi$ to the RHS of Eq. (1a), where $H_{max}$ is
given by Eq. (2). Using the stability analysis of the linearized
Ginzburg-Landau equations near a specific metastable state as
presented in Ref. \cite{our2} we obtain the modified critical
momentum
\begin{equation} 
p_c=\frac{1}{\sqrt{3}}\sqrt{1-\left(\frac{H}{H_{max}}\right)^2+\frac{1}{2R^2}}.
\end{equation}
Note that now $p_c$ decreases with increasing magnetic field. This
automatically leads to a decreasing value of the jump in the
vorticity $\Delta L$ at high magnetic field, because in Ref.
\cite{our1} it was shown that
\begin{equation} 
\Delta L_{max}={\rm Nint}(p_cR),
\end{equation}
where ${\rm Nint}(x)$ returns the nearest integer to the argument.

In order to check the validity of Eq. (6) we performed a numerical
simulation of the two-dimensional Ginzburg-Landau equations, Eqs.
(1a,b), for a ring with $R=5.5\xi$ and $w=1.5\xi$ (for these
parameters the theoretical findings fit the experimental results -
see section below). In Fig. 2 the magnetization, the order
parameter and the gauge-invariant momentum $p$ are shown as
function of the applied magnetic field. The magnetic field was
cycled up and down from $H<-H_{max}$ to $H>H_{max}$. The condition
(6) leads to an hysteresis of $M(H)$ and to a changing value of
the jump in the vorticity in accordance with the change in $p_c$.
The main difference between our theoretical prediction (6) and the
results of our numerical calculations appears at fields close to
$H_{max}$. Apparently it is connected with the fact that for the
considered ring the distribution of the order parameter along the
width of the ring is appreciably nonuniform at $H \simeq H_{max}$
and as a consequence the one-dimensional model breaks down (see
Fig. 1).
\begin{figure}[h]
\includegraphics[width=0.5\textwidth]{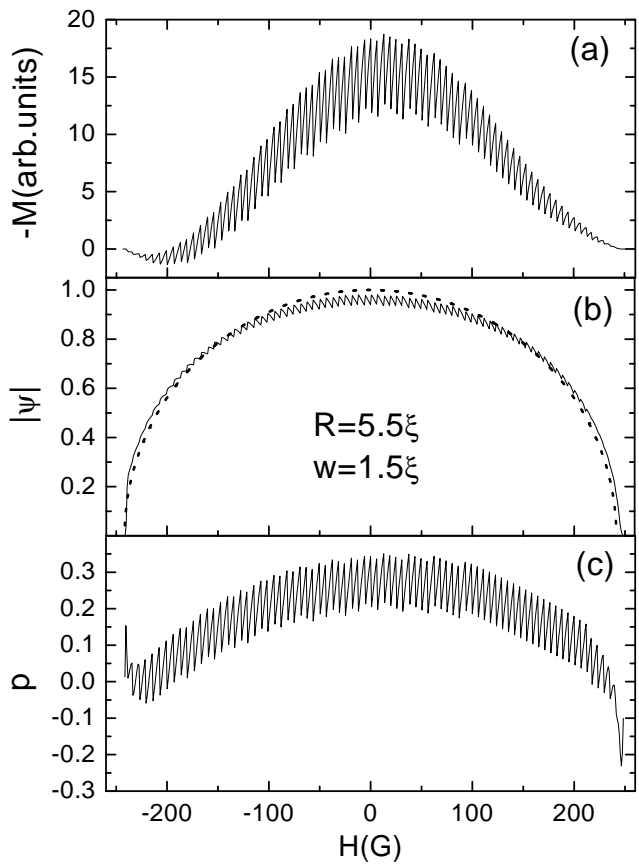}
\caption{Magnetic field dependence of the magnetization (a), the
order parameter (b) and the gauge invariant momentum (c) (in the
middle of the ring) of a ring containing one defect. Dotted line
in Fig. 3(b) is the expression $\sqrt{1-(H/H_{max})^2}$.}
\end{figure}

Finally, we also considered the same ring with a defect. The
effect of the defect was modelled by introducing in the RHS of Eq.
(1a) the term $-\rho(s)\psi$ ($s$ is the arc-coordinate) where
$\rho(s)=-1$ inside the defect region with size $\xi$ and
$\rho(s)=0$ outside. This leads to the results shown in Fig. 3 for
$M(H)$, $|\psi|(H)$ and $p(H)$. Due to the presence of the defect,
$p_c$ differs from Eq. (6) already at low magnetic field
($p_c(H=0) \simeq 0.33$ at given "strength" of the defect) and as
a result only jumps with $\Delta L=1$ are possible in such a ring.
In this case the $p_c$ and $|\psi|$ also depend on the applied
magnetic field with practically the same functional dependence on
$H$ as Eq. (6).

\section{Comparison with experiment}

The measurements were performed on individual Al superconducting
rings by using ballistic Hall micromagnetometry
\cite{Geim1,Geim2}. The techniques employs small Hall probes
microfabricated from a high-mobility two-dimensional electron gas
(2DEG). The rings - having radii $R \simeq 1\mu m$ and width $w$
ranging from $0.1$ to $0.3\mu m$ - were placed directly on top of
the microfabricated Hall crosses, which had approximately the same
width $b$ of about $2\mu m$ (see micrograph in Fig. 4 for a ring
with an artificial defect). These experimental structures were
prepared by multi-stage electron-beam lithography with the
accuracy of alignment between the stages better than $100 nm$. The
rings studied in this work were thermally evaporated and exhibited
a superconducting transition at about 1.25K. The superconducting
coherence length was $\xi(T=0) \simeq 0.18 \mu m$. The latter was
calculated from the electron mean free path $l \simeq 25nm$ of
macroscopic Al films evaporated simultaneously with the Al rings.
The Hall response, $R_{xy}$, of a ballistic cross is given by the
amount of magnetic flux $\int Bds$ through the central square area
$b\times b$ of the cross \cite{Geim1,Peeters}. For simplicity, one
can view the ballistic magnetometer as an analogue of a
micro-SQUID, which would have a square pick-up loop of size $b$
and superconducting rings placed in its centre. We present our
experimental data in terms of the area magnetization $M=<B>-H$
which is the difference between the applied field $H$ and the
measured field $<B> \sim R_{xy}$. Previously, we have studied
individual superconducting and ferromagnetic disks and found
excellent agreement with the above formula \cite{Geim2,Novoselov}.
For further details about the technique, we refer to our earlier
work \cite{Geim1,Geim2,Peeters}.

\begin{figure}[h]
\includegraphics[width=0.3\textwidth]{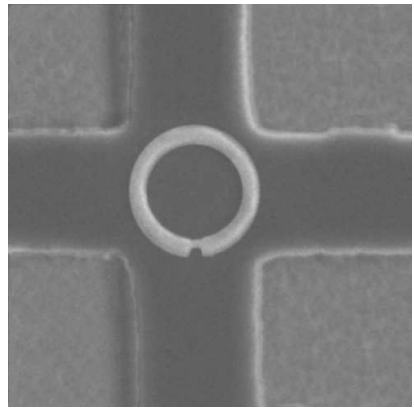}
\caption{A micrograph of the superconducting ring placed on top of
a Hall bar. An artificial defect (narrowing of the ring cross
section) is intentionally made by electron beam lithography.}
\end{figure}

Rings with and without an artificial defect were studied. Let us
consider first the ring without artificial defect. In Fig. 5(a)
the full magnetization loop of such a ring with parameters
$R=1.0\pm 0.1 \mu m$ and $w=0.25\pm 0.05 \mu m $ is shown. In Fig.
6 (solid curve) the low field part of the virgin curve is
presented. From the virgin trace $M(H)$ we can find the magnetic
field for the first vortex entry, $H_{en}$, and hence we estimate
$\xi \simeq 0.19 \mu m$ at the given temperature ($T\simeq 0.4 K$)
using Eq. (5) (this value of $\xi$ is in agreement with the above
experimental value $\xi(0)\simeq 0.18 \mu m$ obtained from the
mean free path). Furthermore, we know from Fig. 5(a) that the
vorticity changes with $\Delta L=3$ for $H\simeq 0$. This agrees
with the fact that the radii of the ring is larger than $4.6\xi$
(see Eq. (7)). Another important information which may be
extracted from the virgin curve is that at the first vortex entry
the magnetization drops considerably but it does not change sign.
If we recall that at every vortex entry $p$ decreases on $1/R$
(and hence the current density $j$ and $M\sim \int [{\bf j} \times
{\bf r}] dV$ also changes proportionally) we can conclude that the
radii of our ring should be in the range $5.5 \xi \lesssim R
\lesssim 6.5 \xi$. This agrees with the experimental value $R/\xi
\simeq 5.3\pm 0.5$.
\begin{figure}[h]
\includegraphics[width=0.5\textwidth]{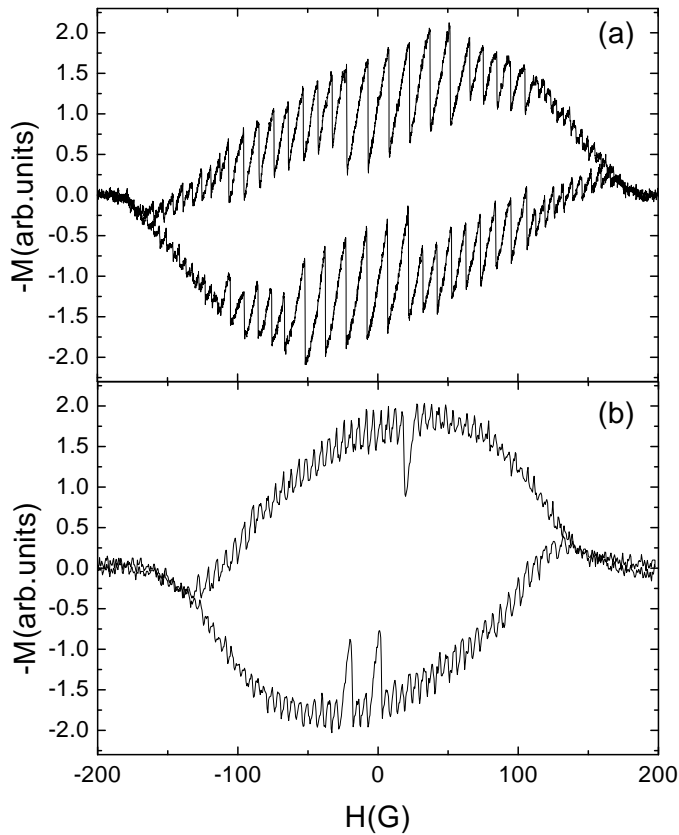}
\caption{Magnetic field dependence of the magnetization of the
ring without (a) and with (b) an artificial defect at $T\simeq 0.4
K$. Parameters of the rings (width and radii) are the same within
experimental accuracy.}
\end{figure}

If we take the above value for $\xi$ and $w \simeq 1.5 \xi$ we
obtain the maximum field of $H_{max}\simeq 223$ G. This value is
slightly smaller than the value obtained from Figs. 2 and 3
$H_{max} \simeq 240 G $ which we attribute to the large coordinate
step which we used in our numerical calculations of Eq. 1(a,b).
The value is also larger than the experimental value
$H_{max}\simeq 185$ G. This disagreement between theory and
experiment is most likely connected to the semi-quantitative
applicability of the Ginzburg-Landau equations in the considered
temperature range. The range of applicability of the
Ginzburg-Landau equations (even stationary ones) for this specific
superconductor is very narrow. Nevertheless based on previous
comparison between experiments and theory for mesoscopic
superconducting disks \cite{Deo1,Deo2} it was found that the GL
equations provided a rather good description of the
superconducting state even deep inside the $(H,T)$ phase diagram.

Figs. 2(a) and 5(a) are qualitatively very similar. For example
our theory describes: i) the hysteresis; ii) the non-unity of the
vorticity jumps, i.e. $\Delta L=3$ in the low magnetic field
region, $\Delta L=2$ in the intermediate H-region, and $\Delta
H=1$ near $H_{max}$. Theoretically (experimentally) we found 6(5),
13(21), 22(18) jumps with respectively $\Delta L=3$, $2$, $1$; and
iii) the non symmetric magnetization near $\pm H_{max}$ for
magnetic field sweep up and down.

In the ring with approximately the same mean radii and width but
containing an intentionally introduced artificial defect, jumps
with $\Delta L=1$ are mostly observed (see Fig. 5(b)). The reason
is that an artificial defect considerably decreases the critical
value $p_c$ (and hence the field $H_{en}$ - see dotted curve in
Fig. 6). From Figs. 2(c), 3(c) it is clear that the maximum value
$p_c^{id} \simeq 0.54$ for a ring without defect and $p_c^{d}
\simeq 0.35$ for a ring with a defect. The ratio
$p_c^{d}/p_c^{id}\simeq 0.65$ is close to the ratio of the field
of first vortex entry $H_{en}^d/H_{en}^{id} \simeq 0.67$ obtained
from experiment (see Fig. 6). From Fig. 2(c) it is easy to see
that for a ring without a defect at $p\simeq 0.35$ there are only
jumps with $\Delta L =1$. But if we slightly increase $p$ then
jumps with $\Delta L=2$ can appear in the system. So we can
conclude that $p=0.35$ is close to the border value which
separates regimes with jumps in vorticity of $\Delta L=1$ and
$\Delta L=2$. From our experimental data it follows that the
maximum value of $p_c$ is very close to this border. Thermal
fluctuations may influence the value of $\Delta L$, in particular
for a $p_c$ value close to this border value. This is probably the
reason why in the experiment (Fig. 5(b)) occasional jumps with
$\Delta L=2$ are observed which are absent in our simulation (Fig.
3(a)).
\begin{figure}[h]
\includegraphics[width=0.5\textwidth]{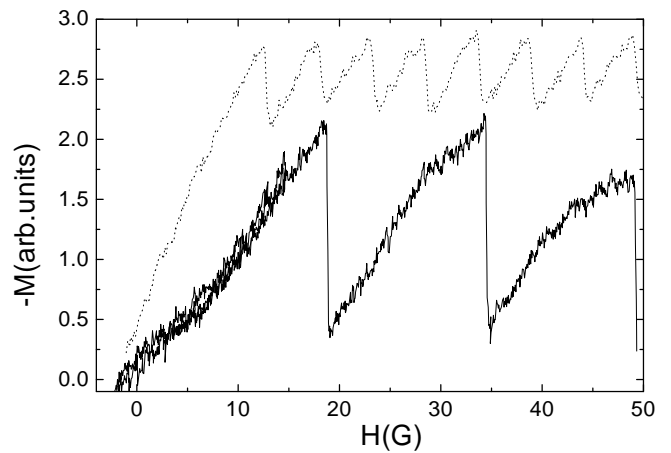}
\caption{Magnetic field dependence of the virgin magnetization of
a ring without (solid curve) and with (dotted curve) an artificial
defect. The dotted curve is shifted for clarity by 0.6 .}
\end{figure}

\section{Conclusion}

We studied multiple flux jumps and irreversible behavior of the
magnetization of thin mesoscopic Al superconducting rings. We have
shown experimentally and theoretically that at low magnetic fields
and for rings with sufficiently large radii the vorticity may
change by values larger than unity. With increasing magnetic field
the order parameter gradually decreases and thus leads to a
decrease of the size of the jumps in the vorticity. For rings with
width less than $2\xi$ analytical expressions were obtained for
the dependence of the order parameter on the applied magnetic
field. We have found that a state with a given vorticity becomes
unstable when the value of the gauge-invariant momentum reaches a
critical value $p_c$ which decreases with increasing magnetic
field. This is responsible for the fact that $\Delta L$ decreases
with increasing $H$. The introduction of an artificial defect in
the ring leads to a decrease of $p_c$ in comparison to the case of
a ring without a defect and also results in a decrease of $\Delta
L$.

\begin{acknowledgments}
This work was supported by the Flemish Science Foundation
(FWO-Vl), the "Onderzoeksraad van de Universiteit Antwerpen", the
"Interuniversity Poles of Attraction Program - Belgian State,
Prime Minister's Office - Federal Office for Scientific, Technical
and Cultural Affairs", EPSRC (UK), and the European ESF-network on
Vortex Matter. One of us (D.Y.V.) was supported by FWO-Vl.
\end{acknowledgments}

\end{document}